\documentclass{epl}

\title{Probing the light induced dipole-dipole interaction in momentum space}
\author{R. L\"OW\inst{1} \and R. GATI\inst{2} \and J. STUHLER\inst{1} \and  T. PFAU\inst{1}}
\institute{
  \inst{1} 5. Physikalisches Intitut, Universit\"at Stuttgart, 70550 Stuttgart\\
  \inst{2} Kirchhoff Institut, Universit\"at Heidelberg, 69120 Heidelberg \\}

\pacs{42.50.Ct}{Quantum description of interaction of light and
matter} \pacs{34.20.Cf}{Interatomic potentials and forces}
\pacs{03.75.-b}{Matter waves}

\begin{document}

\maketitle

\begin{abstract}

We theoretically investigate the mechanical effect of the
light-induced dipole-dipole interaction potential on the atoms in a
Bose-Einstein condensate. We present numerical calculations on the
magnitude and shape of the induced potentials for different
experimentally accessible geometries. It is shown that the
mechanical effect can be distinguished from the effect of incoherent
scattering for an experimentally feasible setting.

\end{abstract}

\section{Introduction}

The interaction of a radiation field with an ensemble of atoms has
been investigated in many distinct contexts since the early works of
Lorentz and Lorenz \cite{lorentz} of light in dense media. Within
one century, researchers came up with an innumerable variety of
light-matter interaction types \cite{born} and reached a recent
summit with the development of laser cooling \cite{nobel97}. The
dipole character of an atom, driven by an electromagnetic wave, can
be used to apply forces on atoms by intensity gradients of light
fields \cite{dipoletrap} but it can also produce forces between
atoms \cite{optbind}. The implication of interacting atoms on the
refractive index was treated theoretically for dilute Bose gases
\cite{dalibard} with an extended Lotentz-Lorenz model. The effect of
light-induced or excited state collisions shows up in laser cooling
by losses in magneto-optical traps \cite{dipoleloss}. The coherent
part of this interaction is recently discussed in basically two
directions. The one focuses on the properties of the light absorbed
and emitted in dense media, namely radiation trapping
\cite{radtrap}, level shifts \cite{dipshift}, dipole blockade
effects as a tool for quantum information processing
\cite{dipolezoller} and collective effects such as superradiance
\cite{dicke}. The other main focus tends towards new forms of
interaction in Bose-Einstein condensates as effective 1/r potentials
\cite{gravity} or as a generator for rotons \cite{roton}. Recently,
the effect of the resonant dipole-dipole interaction in a cold cloud
of highly excited Rydberg states was observed by broadening of
spectroscopic lines \cite{martin} and by resonant energy transfer
between different Rydberg states \cite{noel}.

In this letter, we investigate new physical aspects that arise in
dense cold atomic samples irradiated by a near resonant laser beam.
Atoms exposed to an electromagnetic wave responds as damped harmonic
oscillators and exhibit an alternating electric dipole moment. The
interaction energy of such dipoles can exceed the one of magnetic
dipoles in atomic ground states by several orders of magnitude. We
propose and theoretically investigate an experiment which allows to
study the coherent interaction of laser induced electric dipoles by
transferring the interaction energy among the dipoles into kinetic
energy, which can be probed with standard time of flight techniques.
Initially, the dipoles are generated for a certain flash time by a
laser beam with linear polarization in a spin polarized sample of
cold atoms. During the flash time, the dipole moments reach a steady
state and the light-induced dipole-dipole interaction potential is
build up. As a first step, we calculate this potentials for a
certain density distribution including the retardation effects of
the dipolar fields and the driving electromagnetic wave. The flash
time is chosen long enough that the atoms can evolve in the induced
potential, namely to gain momentum, but short enough not to change
the initial density distribution. As a next step we discuss the
increase of the initial momentum distribution for different
geometries of the atomic cloud and as a function of the angle of the
linear polarization of the laser light. Finally, the results are
compared with parasite effects like spontaneous scattering.

\section{Outline of the calculation}

In the following, we are dealing with oscillating dipoles driven by
an electromagnetic wave with linear polarization and wave vector
$k$. We assume that all dipole moments are of equal strength,
oriented parallel and oscillate with the same frequency. The phase
between two dipoles depends on the position of the atoms with
respect to the electromagnetic wave phase fronts and the interatomic
distance. The retarded interaction potential for two interacting
dipoles with one dipole located at the origin and the other at
$\vec{r_0}$ reads \cite{craig}

\begin{eqnarray} \label{eq.liddi}
\lefteqn{\tilde{V}_{\textrm{dd}}(\vec{r}_0,0)= \frac{d^2
\cos(\vec{k}\cdot\vec{r_0})}{4\pi\varepsilon_0
r_0^3} \: \cdot} \\
& \cdot \sum_{i,j}[(\delta_{ij}-3\frac{r_{0,i} \;
r_{0,j}}{r_0^2})(\cos(k r_0)+k r_0 \sin(k r_0)) -
(\delta_{ij}-\frac{r_{0,i} \; r_{0,j}}{r_0^2})(k^2 r_0^2\cos(k
r_0))] \nonumber \; \;\;\;\; i,j=x,y,z
\end{eqnarray}

where $d$ is the absolute value of the dipole moment and $\vec{k}$
the wave vector of the driving field. Finally, we want to look at a
system of N pairwise interacting dipoles with a density distribution
$n(\vec{r})$. The superposed potential for a dipole at position
$\vec{r}_0$ is given by

\begin{equation} \label{eq.potconv}
V_{\textrm{dd}}(\vec{r}_0)=\int
\tilde{V}_{\textrm{dd}}(\vec{r}_0,\vec{r}) n(\vec{r}) d^3r \; .
\end{equation}

Replacing $\tilde{V}_{\textrm{dd}}(\vec{r}_0,\vec{r})$ by $d^2
\cos(\vec{k} \cdot (\vec{r_0}-\vec{r}))
\textrm{V}^{\prime}_{\textrm{dd}} (\vec{r_0}-\vec{r})$, it is
possible to rewrite the integral as a convolution

\begin{equation} \label{eq.potconv2}
V_{\textrm{dd}}(\vec{r}_0)=\int d^2 \cos(\vec{k} \cdot
(\vec{r_0}-\vec{r})) \textrm{V}^{\prime}_{\textrm{dd}}
(\vec{r_0}-\vec{r}) n(\vec{r}) d^3r \;
\end{equation}

to which the convolution theorem can be applied. Since there exists
no analytical solution, we discreticize the integral
(\ref{eq.potconv2}) on a simple cubic lattice and use FFT algorithms
for evaluation.

The induced dipole potential $V_{\textrm{dd}}$ changes the momentum
distribution of the atomic cloud. In the following, we assume a
density distribution of a Bose-Einstein condensate with all atoms
having the same phase. The undisturbed wave function can be written
as $\psi_0(\vec{r})=e^{i\varphi(t)} \sqrt{n_0(\vec{r})}$. The wave
function is an eigen-state of the unperturbed situation and
therefore the time evolution operator of the system, after the
interaction is switched on, writes
$\hat{U}(\vec{r},t)=\exp(-iV_{\textrm{dd}}(\vec{r}) t/\hbar)$ and
with this $\psi(\vec{r,t})=\hat{U}(\vec{r},t)\psi_0(\vec{r}) $. Here
we demand that the density distribution does not change during the
interaction time, which is legitime in the so called Raman Nath
regime. The Raman-Nath approximation is valid as long the gained
kinetic energy is much smaller than the interaction potential. The
momentum distribution after the interaction time t is given by

\begin{equation}\label{Uaufpsi}
\tilde{n}(\vec{k},t)=\left| \int e^{i\vec{k}\vec{r}}
\hat{U}(\vec{r},t) \psi_0(\vec{r},0) d^3r \right|^2/8 \pi^3 .
\end{equation}

To evaluate the magnitude of the dipole moment, we consider the
atoms as two level systems driven by a coherent light field. This
system has a steady state dipole moment, which arises from a
superposition of the atomic ground and excited state. The amplitude
of this oscillating dipole can be derived using optical Bloch
equations \cite{alleneberly}. The steady state dipole moment reads

\begin{equation} \label{eq.dsteady}
d=2 \frac{d_{ge}}{\Omega} \frac{s}{s+1} \sqrt{\delta^2+\Gamma^2/4}
\end{equation}

where $d_{ge}$ is the dipole matrix element, $\Omega=\Gamma
\sqrt{I_0/2 I_{\rm sat}}$ the Rabi frequency, $\Gamma$ the natural
linewidth, $\delta$ the laser detuning from resonance and the
saturation parameter

\begin{equation}
s=\frac{I_0}{I_{\rm sat}} \frac{1}{1+4 (\delta/\Gamma)^2} .
\end{equation}

$I_0$ is the intensity of the flashing laser and $I_{\rm sat}=\pi h
c \Gamma /3 \lambda^3$ the saturation intensity. A maximum dipole
moment only depends on atomic parameters

\begin{equation}
d_{\rm max}=\sqrt{\frac{3 \Gamma \varepsilon_0hc^3}{4\omega_0^3}}
\end{equation}

and is reached for any detuning if the intensity is set to
$I_0=I_{\rm sat} (1+4\delta^2/\Gamma^2)$ with $\omega_0$ being the
transition frequency. The maximum dipole moment coincides with a
saturation parameter of one and therefore the spontaneous scattering
rate is fixed to $\Gamma/4$.

\section{Numerical calculations-realistsic experimental setups}

\begin{figure}[width=5cm]
\onefigure[width=7cm]{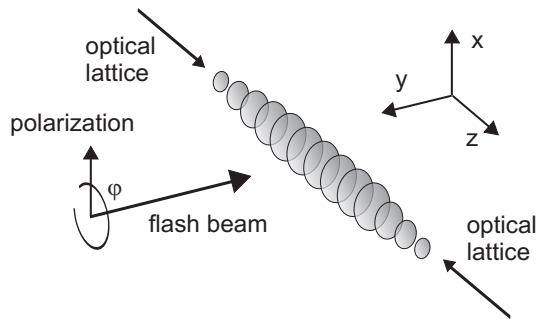} \caption{A Bose-Einstein
condensate within an one-dimensional optical lattice. The resulting
density distribution is a stack of pancake-shaped condensates. A
laser along the y-direction induces the dipole-dipole interaction.
The linear polarization of the flash beam can be altered from
x-polarization ($\varphi$=0$^{\circ}$) to z-polarization
($\varphi$=90$^{\circ}$). The altered momentum distribution is
either projected onto the x-z plane or the y-z plane.} \label{f.1}
\end{figure}

To illustrate the mechanical effect of the dipole-dipole
interaction, we choose a specific example which corresponds to
parameters that are realistic in a typical Bose-Einstein condensate
experiment with Rb$^{87}$ atoms in the F=2, m$_{\rm F}$=2 ground
state. The atomic cloud is confined in a cigar-shaped magnetic trap
with the long axes along z as depicted in fig. \ref{f.1}. In
addition, we adiabatically switch on a retroreflected laser beam at
785nm to create an optical lattice along the z-axis to increase the
density. The depth of the lattice is set to 100 recoil energies,
which results in an axial trapping frequency of 105 kHz. The
additional axial confinement due to the magnetic trap is neglected.
Radially, the trapping frequency is set to 1 kHz. Due to the strong
axial confinement, the density distribution of the ground state can
not be calculated in the Thomas-Fermi approximation. We solved the
full Gross-Pitaevskii equation \cite{stringari} numerically with an
imaginary time Schr\"odinger equation for 250 atoms in a single
lattice site for the given trapping frequencies. The resulting
pancake-shaped density distribution can be approximated in radial
direction by a parabola with a Thomas-Fermi radius of 1.15 $\mu$m
and axially by an Gaussian distribution with a 1/e$^2$ radius of
34.2 nm. The resulting peak density is 9.7 $\times$ 10$^{20}$
m$^{-3}$ and the chemical potential is 5.8 kHz. In the following, we
assume an infinite stack of equal pancakes separated by
$\lambda$/2=785/2 nm. The calculation is carried out initially on a
single pancake and is finally superposed according to the infinite
stack. The grid for the numerical calculation is chosen to be
$64\times 64\times 128=65.536$ lattice points and the grid lattice
spacing is $\lambda_{ol}/32=24.5$ nm where $\lambda_{l}$ is the
wavelength of the laser generating the optical lattice. The
direction of the detuning of the laser affects only slightly the
lattice constant and with this the superposition of the constituent
potentials of each pancake.

The intensity of the flash beam is set to 1120 $I_{\rm sat}$ and its
frequency is detuned 100 MHz from the F=2 to F=3 transition (D2
multiplet of Rubidium) at 780.249 nm, which corresponds to a
detuning of 16.7 $\Gamma$. Using such a large detuning, one can
neglect an inhomogeneous illumination of the atomic cloud since only
a small fraction of the light is absorbed. The induced dipole
potentials alters the effective detuning to the flash beam which
results in an altering phase and magnitude of the oscillating
dipoles within the cloud. This effect can also be neglected, since
the detuning of the flash beam is much larger than the induced
dipole potential. Also nonlinear effects as lensing by the
inhomogeneous density distribution and radiation trapping are
strongly suppressed. The flash beam propagates along the y-direction
and its polarization angle $\varphi$ can be altered from $0^{\circ}$
(polarization along the x-axis) to $90^{\circ}$ (polarization along
the z-axis). The steady state dipole moment with this parameters is
5.26 Debye. The flash time was set to 300 ns, long compared to
$1/\Gamma$, so that all dipoles are in a steady state, but short
enough for being in Raman Nath regime. This means also that
superradiant effects can be neglected \cite{schneble}. The atoms are
initially prepared in the F=2,m$_F$=2 state in respect to the
z-axis.

To account for additional broadening effects of the momentum
distribution by spontaneous scattering events, the time evolution of
the full atomic density matrix was carried out. This includes the
different polarizations of the driving field, the pumping of the
atoms into other m$_F$ states and the angular distribution of the
acquired recoil. For given parameters, about three photons are
scattered per atom. The amount of scattered photons can be
experimentally checked for consistency purposes by the shift of the
center of mass position after some time of flight of the atom cloud.

The final momentum distribution is given by the convolution of the
momentum gain due to the induced dipole potential, the mean field of
the Bose-Einstein condensate and the spontaneously scattered
photons. To extract a mean momentum broadening, the convoluted
distribution was fitted with very good agreement by Gaussian
distribution.

\section{Results}

\begin{figure}
\twoimages[width=7cm]{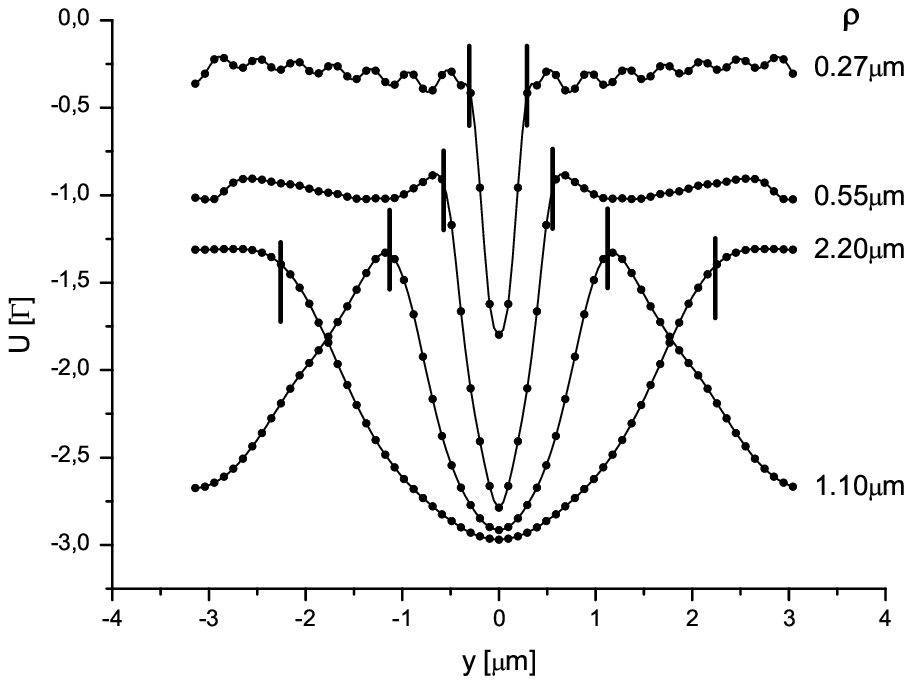}{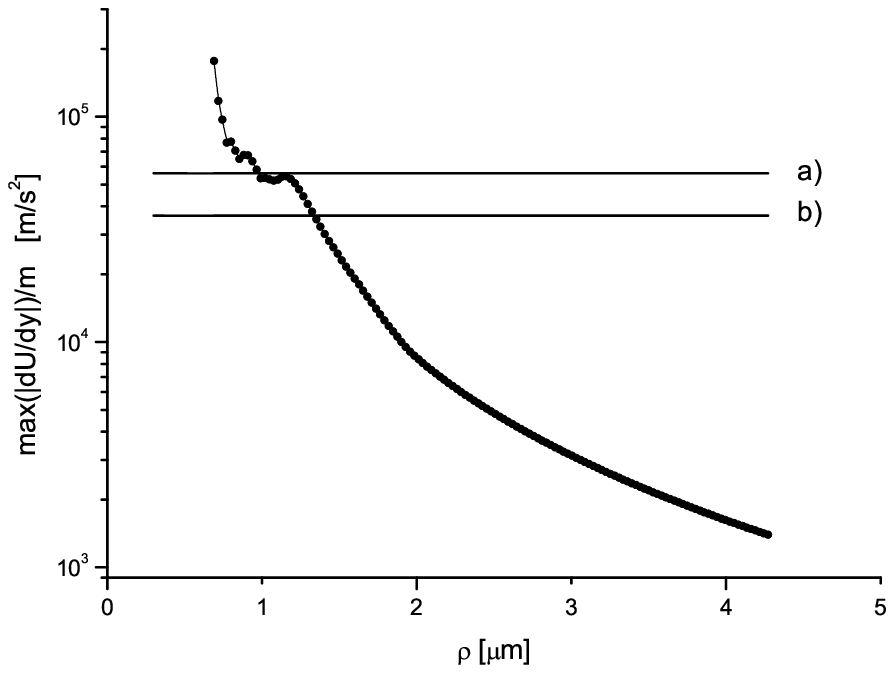} \caption{The
left figure shows the induced potentials along the y-axes at x=z=0
for different radial widths $\rho$ of the atomic clouds. The numbers
and the vertical bars at each plot indicate the radial Thomas-Fermi
radius. The peak atomic density was kept constant at $9.7 \times
10^{20} {\rm m}^{-3}$ for all widths. The polarization of the flash
beam points always along the x-axis. On the right side, the maximum
acceleration as dependence of the radial Thomas-Fermi radius $\rho$
is depicted. The data shows the maximum gradient of the potentials
divided by the mass of a Rubidium atom (m=1.44 $\times 10^{-25}$kg).
The horizontal line a) gives an upper limit for the unidirectional
acceleration ($\frac{d\langle p\rangle}{dt}/m=\hbar k \Gamma /4m$)
due to radiation pressure. The horizontal line b) marks the maximum
acceleration due to momentum diffusion caused by spontaneous
emission processes ($\frac{d \sqrt{\langle p^2 \rangle}}{dt}/m$).
The flashing laser has a detuning of 100 MHz and an intensity of
1120 I$_{\rm sat}$. } \label{f.2}
\end{figure}

\begin{figure}
\twoimages[width=7cm]{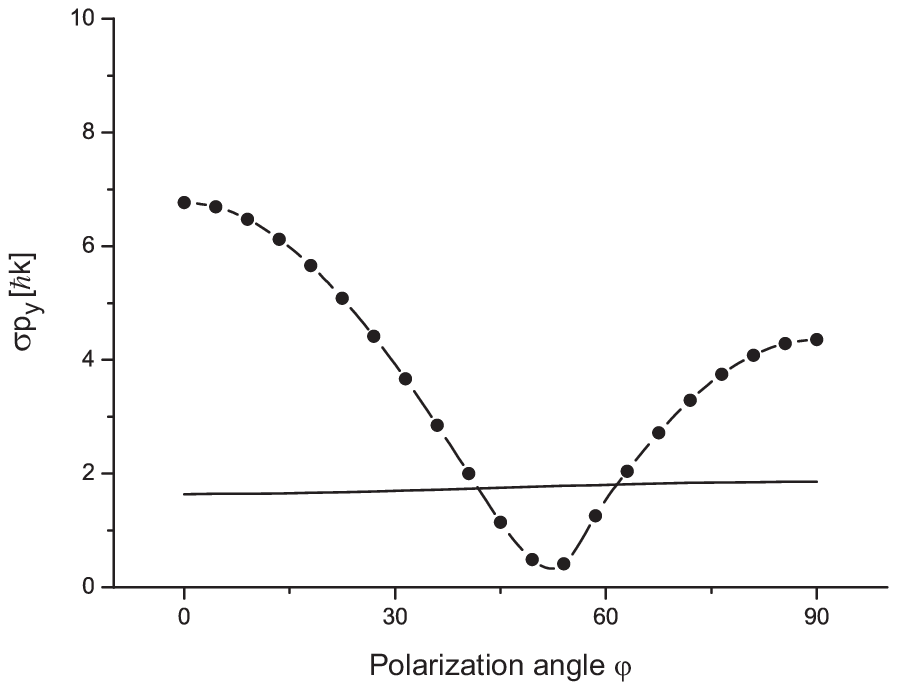}{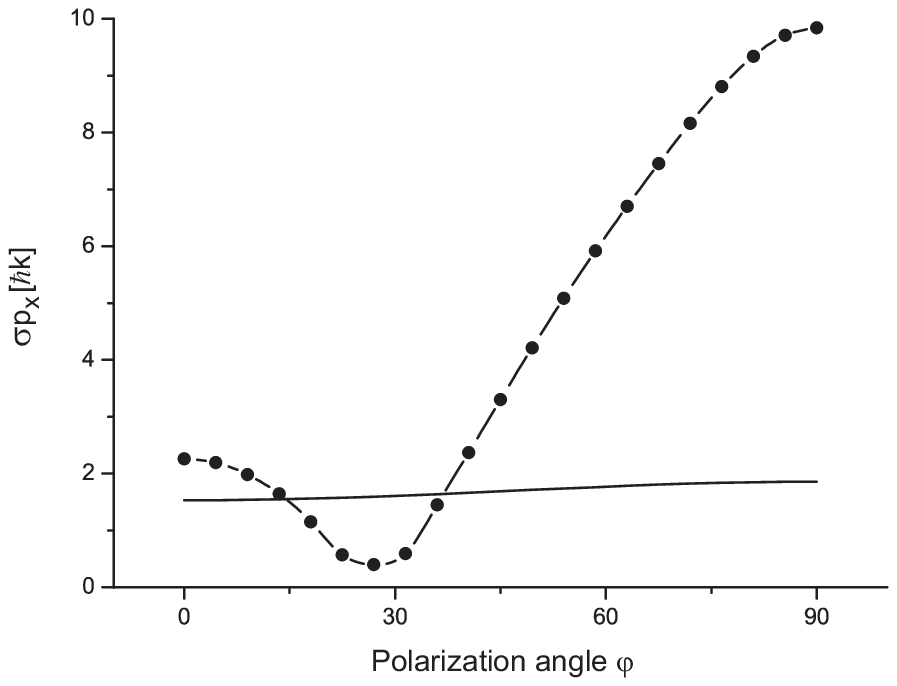} \caption{The two
graphs depict the width of the momentum distribution of the atomic
cloud after the light-induced dipole-dipole potential was applied
for 300ns. The Thomas Fermi radius was set to 1.15 $\mu$m. The graph
on the left shows the broadening in momentum space projected onto
the y-axes as a function of the polarization angle and the right the
projection onto the x-axes. The slash-dotted line includes the
effect of the light-induced potentials and the chemical potential.
The solid line is the incoherent background of the spontaneous
scattered photons as a function of the polarization angle $\varphi$.
To receive the full width in momentum space one has to add the two
curves in quadrature.} \label{f.3}
\end{figure}

In fig. \ref{f.2} the distribution of the light-induced
dipole-dipole interaction potential through the center of a pancake
along the y-axis is shown. The contribution of the neighbouring
pancakes is included by periodic boundary conditions. The different
curves represent pancakes with different radial sizes but at fixed
axial size and constant density and show the dependence of the
induced potentials on the geometry of the atomic cloud. In the limit
of an infinite cloud with a constant density distribution, the
potential would be just a constant and its mechanical effect on the
atoms vanishes. The increase of the induced potentials in the center
of the cloud with larger radii arises form the greater atom number
within the cloud, since the density is kept constant. The
interaction potential is on the order of several MHz, which is large
compared to all other energy scales in the system like the trapping
frequencies and the chemical potential due to the interaction via
s-wave scattering.

The right hand side of fig. \ref{f.2} shows the maximum acceleration
extracted from the potentials. For sufficiently small radial sizes
of the atomic cloud, the maximum acceleration prevails the
unidirectional acceleration $\frac{\hbar k \Gamma}{4}$ due to the
spontaneous light force. This allows to clearly distinguish the
effects emerging from the dipole-dipole interaction from spontaneous
scattering events.

Finally the slash-dotted curves in fig. \ref{f.3} show the
calculated widths of the momentum distribution for the previous
parameters along the x and y direction as a function of the
polarization angle. Not included in the slash-dotted curves is the
contribution of the spontaneous scattering events represented by the
solid lines. The broadening in momentum space can be up to 10
recoils, which is fairly larger than the contribution of the
chemical potential. Noticable is the existence of a strong
dependence of the broadening on the polarization angle $\varphi$. In
both directions exists an angle at which the effect of the
dipole-dipole interaction nearly vanishes and the momentum
distribution is dominated by the released chemical potential. The
plot on the left side shows a minimum close to the so-called magic
angle at 54.74$^{\circ}$ where the interaction of two dipoles
vanishes. Such a minimum broadening is a clear signature of the
dipolar character of the potentials since it can not be explained by
other light-atom interaction mechanisms.

\section{Conclusion}

We have identified a new regime of coherent light-atom interaction,
where novel coherent mechanical effects due to dipole-dipole
interactions are predicted. The calculations show that this
mechanical effect of the light-induced potentials can be detected
experimentally for realistic experimental parameters. The distinct
angular dependence of the polarization of the light field is a clear
indication for a dipole-dipole interaction. The numerical treatment
of the incoherent background of spontaneous scattered photons show
that the signal of the desired mechanical effect is about ten times
larger and exhibits not such a characteristic angular dependence. By
carrying out the proposed experiment, one gains insight in the
physics of coherent dipole-dipole interaction and the feasibility
for its usage for quantum information processing
\cite{dipolezoller}.

We thank M. Lewenstein and L. Santos for helpful discussions and the
DFG for the financial support within the priority program SPP1116.

\end{document}